\documentclass[12pt,conference,onecolumn]{IEEEtran}
\usepackage{times}
\usepackage{epsfig}
\usepackage{amsmath}
\usepackage{amsfonts}
\usepackage{amssymb}
\usepackage{amstext}
\usepackage{latexsym}
\usepackage{color}
\usepackage{ifthen}
\usepackage{multirow}
\usepackage{subfigure}
\newcommand{\be}{\begin{equation}}
\newcommand{\ee}{\end{equation}}
\newcommand{\bear}{\begin{eqnarray}}
\newcommand{\eear}{\end{eqnarray}}
\newcommand{\bears}{\begin{eqnarray*}}
\newcommand{\eears}{\end{eqnarray*}}
\newcommand{\bi}{\begin{itemize}}
\newcommand{\ei}{\end{itemize}}
\newcommand{\ben}{\begin{enumerate}}
\newcommand{\een}{\end{enumerate}}

\newcommand{\beq}{\begin{equation}}
\newcommand{\eeq}{\end{equation}}

\newcommand{\lp}{ \left(}
\newcommand{\rp}{ \right)}

\newtheorem{theorem}{Theorem}[section]

\newtheorem{defn}[theorem]{Definition} 
\newtheorem{lemma}[theorem]{Lemma}

\newcommand{\xbf}{\mbox{${\bf x }$} }
\newcommand{\ybf}{\mbox{${\bf y }$} }

\newcommand{\Xbf}{\mbox{${\bf X }$} }
\newcommand{\Ybf}{\mbox{${\bf Y }$} }

\newcommand{\Prob}{\mbox{${\mathbb P}$} }

\newcommand{\prob}[1]{\Prob \left\{ #1 \right\}}

\begin{document}

\title{Approximate Capacity of Gaussian Relay Networks}

\author{\authorblockN{Amir Salman Avestimehr}
\authorblockA{Wireless Foundations\\
 UC Berkeley,\\
 Berkeley, California, USA.\\
{\sffamily avestime@eecs.berkeley.edu}} \and
\authorblockN{Suhas
N. Diggavi}
\authorblockA{School of Computer and\\
Communication Sciences, EPFL,\\
 Lausanne, Switzerland.\\
 {\sffamily suhas.diggavi@epfl.ch}}
\and
\authorblockN{David N C. Tse}
\authorblockA{ Wireless Foundations\\
 UC Berkeley,\\
 Berkeley, California, USA.\\
 {\sffamily dtse@eecs.berkeley.edu}}
 }

\maketitle

\begin{abstract}
We present an achievable rate for general Gaussian relay networks. We show that the achievable rate is within a constant number of bits from the information-theoretic cut-set upper bound on the capacity of these networks. This constant depends on the topology of the network, but not the values of the channel gains. Therefore, we uniformly characterize the capacity of Gaussian relay networks within a constant number of bits, for all channel parameters.
\end{abstract}

%part of this paper includes:
%1-Introduction
%2-Proof for layered networks
%3-proof for arbitrary networks

\section{Introduction}
\label{sec:intro}

%Consider a network represented by a directed relay network
%$\mathcal{G}=(\mathcal{V},\mathcal{E})$ where $\mathcal{V}$ are the
%vertices representing the communication nodes in the relay network.
%The communication problem considered is unicast (or multicast with all
%destinations requesting the {\em same} message). Therefore a special
%node $S\in\mathcal{V}$ is considered the source of the message and a
%special node $D\in\mathcal{V}$ is the intended destination. All other
%nodes in the network facilitate communication between $S$ and $D$.  

%Since Van der Meulen introduced the relay channel in 1971, there has been an extensive amount of research on determining the capacity of this network, specially the Gaussian relay channel. So far, the only known upper bound on the capacity is the information-theoretic cut-set upper bound. 

%quite a few communication strategies have been developed for the relay channel, a network with only one relay. Most of these schemes can be traced back to the seminal work of Cover and El Gamal \cite{coverRelay}. There has also been some work on generalizing these strategies for relay networks, for example \cite{michaelPaper}. The only known upper bound on the capacity of Gaussian relay networks is the information theoretic cut-set upper bound and none of these schemes are tight for any range of channel parameters. In fact, in a general network with wide range of channel parameters, even the performance of these schemes are not clear.

Characterizing the capacity of wireless relay networks has been a challenging problem over the past couple of decades. Although, many communication schemes have been developed \cite{coverRelay}-\cite{michaelRelay}, the capacity of even the simplest Gaussian relay network: single source, single destination, single relay, is still unknown. In general, the only known upper bound on the capacity of Gaussian relay networks is the information theoretic cut-set upper bound which is not achieved by any of those schemes, not even for a fixed realization of the channel gains. Furthermore, in a general network with a wide range of channel parameters, the gap between those achievable rates and the cut-set upper bound is unclear. As a result, we do not even have a good approximation of the capacity with an explicit guarantee.

In this paper we introduce a simple coding strategy for general Gaussian relay networks. In this scheme each relay first quantizes the received signal at the noise level, then randomly maps it to a Gaussian codeword and transmits it. We show that  we can achieve a rate  that is guranteed to be within a constant gap from the cutset bound.  This constant depends on the topological parameters of the network (number of nodes in the network), but not on the values of the channel gains. Therefore, we get a uniformly good approximation of the capacity of Gaussian relay networks, uniform over all values of the channel gains, thus particularly good approx at high SNR. The presented scheme has close connections to the random coding scheme introduced in \cite{ACLY00} to achieve the capacity of wireline networks. It has also some connections with the compress, hash, and forward protocol described in \cite{kimPaper}, except here the destination is not required to decode the quantized signals at the relays.

The ideas for the main approximation result were inspired by the insight obtained by analyzing deterministic relay networks (see \cite{ADTallerton2_07}). The deterministic approach was motivated by the development of the linear deterministic model (see \cite{ADTitw07}, \cite{ADTallerton1_07}), which was seen to capture the key features of wireless channels. We developed some of the connections between the linear deterministic relay network and the Gaussian relay network in  \cite{ADTallerton1_07}.

\section{Problem statement and main results}
\label{sec:MainRes}

Consider a network represented by a directed relay network
$\mathcal{G}=(\mathcal{V},\mathcal{E})$ where $\mathcal{V}$ is the
set of vertices representing the communication nodes in the relay network, and $\mathcal{E}$ is the set of edges between nodes.
The communication problem considered is unicast. Therefore a special
node $S\in\mathcal{V}$ is considered the source of the message and a
special node $D\in\mathcal{V}$ is the intended destination. All other
nodes in the network facilitate communication between $S$ and $D$.  The received signal
$y_j$ at node $j\in\mathcal{V}$ and time $t$ is given by
\begin{equation}
\label{eq:GenDetModel}
y_{j}^{[t]}=\sum_{i\in \mathcal{N}_j} h_{ij} x_i^{[t]} + z_j^{[t]}\end{equation}
where each $h_{ij}$ is a complex number representing the channel gain from node $i$ to node $j$, and  $\mathcal{N}_j$ is the set of nodes that are neighbors of $j$ in $\mathcal{G}$. Furthermore, we assume there is an average power constraint equal to 1 at each transmitter. Also $z_j$, representing the channel noise, is modeled as as complex normal (Gaussian) random variable
\beq z_j \sim \mathcal{CN} (0,1) \eeq

For any relay network, there is a natural information-theoretic
cut-set bound \cite{CoverThomas91}, which upperbounds the reliable
transmission rate $R$:
\begin{eqnarray}
\label{eq:MinCut}
R &< &\overline{C}=\max_{p(\{x_j\}_{j\in\mathcal{V}})} \min_{\Omega\in\Lambda_D}
I(Y_{\Omega^c};X_{\Omega}|X_{\Omega^c}) 
\end{eqnarray}
where $\Lambda_D=\{\Omega:S\in\Omega,D\in\Omega^c\}$ is all
source-destination cuts (partitions).

The following is our main result
\begin{theorem}
\label{thm:Main}
Given a Gaussian relay network, $\mathcal{G}=(\mathcal{V},\mathcal{E})$, we can achieve all rates  $R$ up to $\overline{C}-\kappa$. Therefore the capacity of this network satisfies
\begin{equation}
\label{eq:Main}
\overline{C}-\kappa \leq C \leq \overline{C}
\end{equation}
Where $\overline{C}$ is the cut-set upper bound on the capacity of $\mathcal{G}$ as described in equation (\ref{eq:MinCut}), and $\kappa$ is a constant and is upper bounded by $5|V|$, where $|V|$ is the total number of nodes in $\mathcal{G}$.
\end{theorem}

The gap ($\kappa$) holds for all values of the channel gains and is relevant particularly when the SNR is high and the capacity is large. While it is possible to improve $\kappa$ further, in this paper we focus to prove such a constant, depending only on the topology of $\mathcal{G}$ but not the channel parameters, exists in general. This constant gap result is a far stronger result than the degree of freedom result, not only because it is non-asymptotic but also because it is uniform in the many channel SNR's. This is also the first constant gap approximation of the capacity of Gaussian relay networks. As we will discuss in the next section, the gap between the achievable rate of other well known relaying schemes and the cut-set upper bound in general depends on the channel parameters and can become arbitrarily large. 

%Therefore, we have uniformly good approximation of the capacities of Gaussian relay networks, uniform over all values of the channel gains.

%To make the proofs more readable the result for layered networks are proved first for linear %%deterministic model that is more intuitive in section \ref{sec:LayFF} and then for  general %%deterministic model in section \ref{sec:GenDet}. Finally we will discuss time-expansion %%ideas and prove the results for arbitrary networks in section \ref{subsec:TimExp}.

\subsection{Examples}
\label{subsec:OtherSchemes}
In this section we use a few examples to show that the gap between the achievable rate of other relaying schemes and the cut-set upper bound depends on the channel parameters and can become arbitrarily large. In particular we focus on three well known strategies: amplify-forward, decode-forward, and compress-forward.

\subsubsection{Amplify-forward strategy}
\label{subsubsec:AF}
\begin{figure*}[htp]
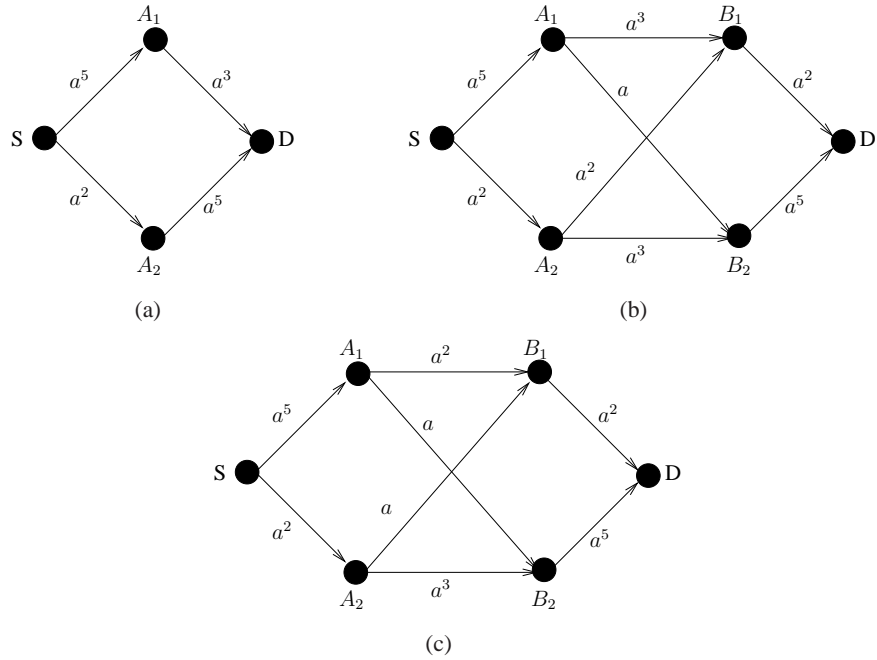

     \centering \subfigure[ ]{
       \scalebox{0.7}{\input{diamond.pstex_t}}
}
     \hspace{0.3in}
     \subfigure[ ]{
       \scalebox{0.7}{ \input{twoLayer.pstex_t}}
} 
\hspace{0.3in}
     \subfigure[ ]{
       \scalebox{0.7}{ \input{twoLayerQ.pstex_t}}
} 
\caption{\label{fig:examples} Diamond network is shown in (a). A two layer network is shown in (b). The effective network for compress-forward strategy is shown in (c).}
\end{figure*}

Consider the diamond network with real channel gains shown in figure \ref{fig:examples}(a). Assume $a$ is a large real number. The cut-set upper bound is approximately,
\beq \label{eq:CutSetDiamond} \overline{C} \approx 5 \log a\eeq

Now consider an amplify-forward strategy in which nodes $A_1$ and $A_2$ amplify the received signal by $\alpha_1$ and $\alpha_2$ and forward them to the destination. Then assuming that $x$ was transmitted at the source, the received signal at the destination will be
\beq \label{eq:AF} y_D=a^3 \alpha_1\lp a^5x+z_{A_1} \rp + a^5 \alpha_2\lp a^2x+z_{A_2} \rp +z_{D} \eeq 
where $z_{A_1}$, $z_{A_2}$ and $z_{D}$ are Gaussian noises with variance 1 and $x$ is the transmitted signal with average power constraint equal to 1. To satisfy the average transmit power constraint at $A_1$ and $A_2$, for large values of $a$ we should have
\beq
\alpha_1 \leq \frac{1}{a^5} , \quad
\alpha_2 \leq \frac{1}{a^2}
\eeq

Now since (\ref{eq:AF}) is just like a point to point channel from $S$ to $D$, the achievable rate of amplify-forward strategy will approximately be
\begin{eqnarray} R_{AF} &=& \frac{1}{2} \log \frac{a^{16} \alpha_1^2+a^{14} \alpha_2^2}{a^6 \alpha_1^2+a^{10} \alpha_2^2+1}\\
& \leq & \frac{1}{2} \log  \frac{2 \max\{a^{16} \alpha_1^2, a^{14} \alpha_2^2\}}{\max\{a^6 \alpha_1^2,a^{10} \alpha_2^2,1\}}    \\
%& \leq & \frac{1}{2} \lp 1+ \max \{ \log a^{16} \alpha_1^2 , \log \frac{a^{14} \alpha_2^2}{a^{10} \alpha_2^2 } \} \rp \\ 
\label{eq:AFUpperBound} & \leq & \frac{1}{2} \lp 1+ 6 \log a \rp    \end{eqnarray} 
Now by comparing (\ref{eq:AFUpperBound}) and (\ref{eq:CutSetDiamond}) we note that as $a$ increases the gap between the achievable rate of amplify-forward strategy and the cut-set upper bound increases. Now by theorem \ref{thm:MainLay} in section \ref{subsec:ProofLay}, which is a special case of our main theorem \ref{thm:Main} for multi-stage networks, the achievable rate of the relaying strategy proposed in this paper is within $\frac{1}{2}12=6$ bits of the cut-set upper bound of this network for all channel parameters\footnote{\label{foot1} factor of $\frac{1}{2}$ comes from the fact that here we are dealing with channels with real valued gains}.

%\begin{figure}
%\centering
%\scalebox{0.6} {\input{diamond.pstex_t}}
%\caption{\label{fig:Diamond} Diamond network}
%\end{figure}

\subsubsection{Decode-forward strategy}
\label{subsubsec:DF}
Consider the same example as shown in figure \ref{fig:examples}(a). Now it is easy to show that the achievable rate of the decode-forward strategy is upper bounded by
\beq \label{eq:DFUpperBound} R_{DF} \leq 3 \log a \eeq Therefore, as $a$ gets larger, the gap between the achievable rate of decode-forward strategy and the cut-set upper bound (\ref{eq:CutSetDiamond}) increases.

\subsubsection{Compress-forward strategy}
\label{subsubsec:CF}
Consider the example shown in figure \ref{fig:examples}(b). For large values of $a$, cut-set upper bound on the capacity of this relay network is approximately
\beq \label{eq:CutSetTwoLayer}  \overline{C} \approx 5 \log a \eeq 

Now consider the compress-forward strategy as described in \cite{michaelRelay} section V. The achievable rate of this scheme is characterized in Theorem 3 (\cite{michaelRelay} page 9), which is in the form of a mutual information maximization over auxiliary random variables $U_{\mathcal{T}}$ and $\hat{Y}_{\mathcal{T}}$. Even though this is written in single-letter form, since there is no cardinality bounds, the rate optimization is still an infinite dimensional optimization problem. However, to simplify this problem further, assume that auxiliary random variables $U_{\mathcal{T}}$ are set to zero, and $\hat{Y}_{\mathcal{T}}$ are restricted to have a Gaussian distribution, which leads to a finite dimensional problem.

The scheme is such that the Wyner$-$Ziv source-coding region  of each layer must intersect the channel-coding region of the next layer. As a result by looking at layer $\{B_1,B_2\}$ we note that node $B_1$ should compress its received signal to a Gaussian random variable with variance $a^2$. In another words, just quantize the received signal with distortion $a$. Therefore the effective network will look like the one shown in figure \ref{fig:examples} (c). Note that now the cut-set upper bound of this network is approximately,  $\overline{C}'  \approx 4 \log a $.

As a result, with this compress-forward scheme, it is not possible to get a rate more than $4 \log a$.  As $a$ increases the gap between the achievable rate of compress-forward strategy and the cut-set upper bound increases. Now by Theorem \ref{thm:MainLay} in section \ref{subsec:ProofLay}, which is a special case of our main Theorem \ref{thm:Main} for multi-stage networks, the achievable rate of the relaying strategy proposed in this paper is within $\frac{1}{2} \times 18=9$ bits of the cut-set upper bound of this network for all channel parameters.

\subsection{Proof Strategy}
\label{subsec:proofStrategy}
Theorem \ref{thm:Main} is the main result of the paper and the
rest of the paper is devoted to sketch its proof. For details of the proof, the reader is referred to \cite{ADTfinal}. First we focus on networks
that have a layered structure, i.e. all paths from the source to the
destination have equal lengths. With this special structure we get a
major simplification: a sequence of messages can each be encoded into
a block of symbols and the blocks do not interact with each other as
they pass through the relay nodes in the network. The proof of the
result for layered network is done in section \ref{sec:Lay}. Second, we
extend the result to an arbitrary network by considering its
time-expanded representation. This is done in section \ref{sec:Gen}\footnote{The concept of time-expanded representation is also used in \cite{ACLY00}, but the use there is to
handle cycles. Our main use is to handle interaction between messages
transmitted at different times, an issue that only arises when there
is interference at nodes.}. The time-expanded network is layered
and we can apply our result in the first step to it. To complete the
proof of the result, we need to establish a connection between the cut
values of the time-expanded network and those of the original
network. We do this using sub-modularity properties of entropy function.
%in Section \ref{subsec:TimExp}.

%\section{Proof of the main result}
%\label{sec:proof}
%In this section we prove main theorem \ref{thm:Main}. We first describe the encoding scheme.

\section{Layered networks}
\label{sec:Lay}
In this section we prove main theorem \ref{thm:Main} for a special case of layered networks, where all paths from the source to the destination in $\mathcal{G}$ have equal length.
In a layered network, for each node $j$ we have a length $l_j$ from the source and
all the incoming signals to node $j$ are from nodes $i$ whose distance
from the source are $l_i=l_j-1$. Therefore, as in the example network
of Figure \ref{fig:Confusability}, we see that there is message
synchronization, {\em i.e.,} all signals arriving at node $j$ are
encoding the same sub-message.

Suppose message $w_k$ is sent by the source in block $k$, then since
each relay $j$ operates only on block of lengths $T$, the signals
received at block $k$ at any relay pertain to only message $w_{k-l_j}$
where $l_j$ is the path length from source to relay $j$.  To
explicitly indicate this we denote by $\ybf_j^{(k)}(w_{k-l_j})$ as the received signal at block $k$ at node $j$. We
also denote the transmitted signal at block $k$ as
$\xbf_j^{(k)}(w_{k-1-l_j})$ .
%which is obtained by
%randomly mapping $\hat{\ybf}_j^{(k-1)}(w_{k-1-l_j})$.

\subsection{Encoding}
\label{subsec:EncLay}
We have a single source $S$ with message $W\in\{1,2,\ldots,2^{RT}\}$
which is encoded by the source $S$ into a signal over $T$
transmission times (symbols), giving an overall transmission rate of
$R$. 

Each relay operates over blocks of time $T$ symbols.  In particular block $k$ of $T$ received
symbols at node $i$ is denoted by
$\ybf_i^{(k)}=\{y_i^{[(k-1)T+1]},\ldots,y_i^{[kT]}\}$ and the transmit
symbols by $\xbf_i^{(k)}$. Now the achievability strategy is the following: each received sequence $\ybf_i^{(k)}$ at node $i$ is quantized into $\hat{\ybf}_i^{(k)}$ which is then randomly mapped  into a Gaussian codeword $\xbf_i^{(k)}$ using a random (binning) function $f_i(\hat{ \ybf}_i^{(k)})$. For quantization, we use a Gaussian vector quantizer.

Since we have a layered network, without loss of generality
consider the message $w=w_1$ transmitted by the source at block
$k=1$. At node $j$ the signals pertaining to this message are received
by the relays at block $l_j$. Given the knowledge of all the encoding functions
 at the relays and signals received at block $l_D$, the decoder $D$, attempts to
decode the message $W$ by finding the message that is jointly typical with its observations.

\subsection{Proof illustration}
\label{subsec:PfIdea}

Consider the encoding-decoding strategy as described in section \ref{subsec:EncLay}. Our goal is to show that, using this strategy, all rates described in the theorem are achievable. The method we use is based on a \emph{distinguishability} argument. This argument was used in \cite{ACLY00} in the case of wireline networks. In \cite{ADTallerton2_07}, we used similar arguments to characterize the capacity of a general class of linear deterministic relay networks with broadcast and multiple access. The main idea behind this approach is the following: due to the deterministic nature of these channels, each message is mapped to a deterministic sequence of transmit codewords through the network. The destination can not distinguish between two messages if and only if its received signal under these two messages are identical. If so, there would be a partition of nodes in the network such that the nodes on one side of the cut can distinguish between these two messages and the rest can not. This naturally corresponds to a cut separating the source and the destination in the network and the probability that this happens can be related to the cut-value. This is the main tool that we used in  \cite{ADTallerton2_07} to show that thecut-set upper bound can actually be achieved.

However, in the noisy case, the difference from the previous
analyses is that each message is potentially mapped to a {\em set} of
possible transmit sequences.  The particular transmit sequence chosen
depends on the noise realization, which can be considered
``typical''. Pictorially it means that there is some fuzziness around the sequence of transmit codewords associated with each message. Hence, two messages will still be distinguishable at the destination if the fuzzy received signal associated with them are not overlapping. This intuitively means that if we can somehow bound this randomness, a communicate rate close to the cut-set bound is achievable.

%Therefore, for the decoder to check if a received
%sequence is jointly typical with a particular message, it would need
%to check through {\em all} members of this set. Hence, two messages will be confused at the destination if the received signal is a member of both typical sets corresponding to the messages. Intuitively, as long as the list sizes corresponding to messages are not growing 

%
%{\em Note that in the
%relay network, only the destination has to do this check, since the
%intermediate relay nodes just do the mapping and are not interested in
%disambiguating the messages.} We only do the distinguishablity at the
%relay nodes as an analysis tool.

\begin{figure}[h]
\centering
\input{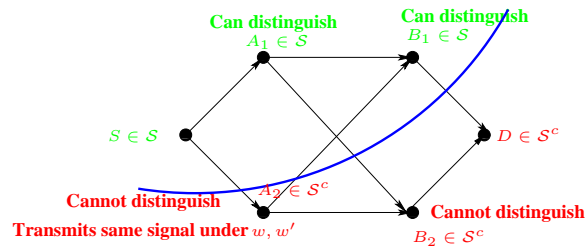}
\caption{An example of a layered Gaussian relay netowrk. }
\label{fig:Confusability}
\end{figure}

In order to illustrate the proof ideas of Theorem
(\ref{thm:Main}) we examine the network shown in Figure
\ref{fig:Confusability}.

% We will analyze this network first for linear
%deterministic model and then we use the same example to illustrate the
%ideas for general deterministic functions in Section
%\ref{subsec:PfIdeaGen}.

Assume a message $w$ is transmitted by the source. Once the destination receives $\ybf_D$, quantizes it to get $\hat{\ybf}_D$. Then, it will decode the message by finding the unique message that is jointly typical with $\hat{\ybf}_D$ (the precise definition of typicality will be given later). An error occurs if either $w$ is not jointly typical with $\hat{\ybf}_D$ or there is another message $w'$ such that $\hat{\ybf}_D$ is jointly typical with {\em both} $w,w'$.

Now for the relay network, a natural way to define whether a
message $w$ is typical with a received sequence is whether we have a
``plausible'' transmit sequence\footnote{Plausibility essentially
means that the transmit sequence is a member of the typical set of
possible transmit sequences under $w$.} under $w$ which is jointly typical
with the received sequence. More formally, we have the following definitions.
\begin{defn}
\label{def:TypRxSet}
For a message $w$, we define the set of received sequences that are typical
with the message as,
\begin{equation}
\label{eq:TypRxSet}
\mathcal{Y}_i(w) = \{\hat{\ybf}_i: (\hat{\ybf}_i,w) \in T_{\delta}\},
\end{equation}
where we still need to define what we mean by $(\hat{\ybf}_i,w) \in
T_{\delta}$.
\end{defn}
\begin{defn}
\label{def:TypTxSet}
For a message $w$, we define the set of transmitted sequences that are
typical with the message as,
\begin{equation}
\label{eq:TypTxSet}
\mathcal{X}_i(w) = \{\xbf_i:
\xbf_i=f_i(\hat{\ybf}_i), \hat{\ybf}_i\in \mathcal{Y}_i(w)\},
\end{equation}
which defines the ``typical'' transmit set associated with a message $w$.
\end{defn}
Note here that since $\xbf_i=f_i(\hat{\ybf}_i)$, then naturally $(\xbf_i,\hat{\ybf}_i) \in T_{\delta}$.
This leads us to the following definition,
\begin{defn}
\label{def:JointTypMsg}
We define $(\hat{\ybf}_i,w) \in T_{\delta}$ if
\begin{equation}
\label{eq:JointTypMsg}
(\hat{\ybf}_i,\{\xbf_j\}_{j\in In(i)}) \in T_{\delta} \mbox{  for some  }
\xbf_j\in\mathcal{X}_j(w), \,\, \forall j\in In(i)
\end{equation}
where $In(i)$ is defined as the set of nodes with signals incident on
node $i$.
\end{defn}

Therefore by this definition, if a message $w$ is typical with a received sequence, we have a
sequence of typical transmit sequences in the network that are jointly typical
with the $w$ and the received sequence at the destination.

Now note the following important observation,
\paragraph*{Observation} Note that if node $i$ cannot distinguish between two
messages $w,w'$, this means that the signal received at node $i$,
$\hat{\ybf}_i$ is such that $(\hat{\ybf}_i,w) \in T_{\delta}$
and $(\hat{\ybf}_i,w') \in T_{\delta}$. Therefore we see that
\begin{equation}
\label{eq:RxInt}
\hat{\ybf}_i \in \mathcal{Y}_i(w) \cap \mathcal{Y}_i(w').
\end{equation}
Due to the mapping $\xbf_i=f_i(\hat{\ybf}_i)$, we therefore see that
$ \xbf_i \in \mathcal{X}_i(w) \cap \mathcal{X}_i(w') $. Therefore, there exists a sequence under $w'$ which is the same as that transmitted under $w$ and could therefore have been
potentially transmitted under $w'$. 

Now, assuming a message $w$ is transmitted by the source, an error occurs at the destination if either $w$ is not jointly typical with $\hat{\ybf}_D$, or there is another message $w'$ such that $\hat{\ybf}_D$ is jointly typical with {\em both} $w,w'$. By the law of large numbers, the probability of the first event becomes arbitrarily small as communication block length, $T$, goes to infinity. So we just need to analyze the probability of the second event. To do so, we evaluate the probability that $\hat{\ybf}_D$ is jointly
typical with both $w$ and $w'$, where $w'$ is another message independent of $w$. Then we use union bound over all $w'$'s to bound the probability of the second event. 

Based on our earlier observation, if  $\hat{\ybf}_D$ is jointly
typical with $w,w'$, then there must be a  typical transmit sequence $\xbf_{\mathcal{V}}'=(\xbf_{S}',\xbf_{A_1}',\xbf_{A_2}',\xbf_{B_1}',\xbf_{B_2}')$ under $w'$ such that,
 $ (\hat{\Ybf}_{D},\xbf_{B_1}',\xbf_{B_2}') \in T_{\delta} $. This means that the destination thinks this is a plausible sequence. Now for any such sequence there is a natural cut, $\Omega$, in $\mathcal{G}$ such that the nodes on the right hand side of the cut ({\em i.e.}  in $\Omega$) can tell $\xbf_{\mathcal{V}}'$ is not a plausible sequence, and those on the left hand side of the cut ({\em i.e.}  in $\Omega^c$) can not.  Clearly this cut is a source-destination partition.

For now, assume that the cut is $\Omega=\{S,A_1,B_1\}$, as shown in figure \ref{fig:Confusability}.
Since $A_2$, $B_2$ and $D$ think $\xbf_{\mathcal{V}}'$ is a plausible sequence, we have
\begin{eqnarray} 
\label{eq:ExmCond1}(\hat{\Ybf}_{A_2},\xbf_{S}' ) & \in & T_{\delta} \\
\label{eq:ExmCond2} (\hat{\Ybf}_{B_2},\xbf_{A_1}',\xbf_{A_2}') & \in & T_{\delta} \\
\label{eq:ExmCond3} (\hat{\Ybf}_{D},\xbf_{B_1}',\xbf_{B_2}') & \in & T_{\delta} 
\end{eqnarray}
For any such sequence $\xbf_{\mathcal{V}}'$, since $w$ is independent of $w'$, we have
\beq \prob{( \hat{\Ybf}_{A_2},\xbf_{S}')  \in  T_{\delta}} \leq 2^{-TI(X_S;Y_{A_2})} \eeq
Now, for the layer $(A_1,A_2)$, we condition
on a particular sequence $\xbf_{A_2}$ to have been transmitted by $A_2$. If  $\xbf'_{A_2}=\xbf_{A_2}$, since  $\xbf'_{A_1}$ is chosen independent of $\xbf_{A_1}$ we have,
\begin{equation}
\label{eq:TypT1}
\prob{(\hat{\Ybf}_{B_2},\xbf'_{A_1},\xbf'_{A_2})\in
T_{\delta}} \leq 2^{-TI(\hat{Y}_{B_2};X_{A_1}|X_{A_2})},
\end{equation}
and similarly  If  $\xbf'_{A_2}\neq \xbf_{A_2}$, since  $\xbf'_{A_1},\xbf'_{A_2}$ are chosen independent of
$\xbf_{A_1},\xbf_{A_2}$ we have,
\begin{eqnarray}
\label{eq:TypT2}
\prob{(\hat{\Ybf}_{B_2},\xbf'_{A_1},\xbf'_{A_2})\in
T_{\delta}} & \leq & 2^{-TI(\hat{Y}_{B_2};X_{A_1},X_{A_2})} \\
& \leq & 2^{-TI(\hat{Y}_{B_2};X_{A_1}|X_{A_2})}
\end{eqnarray}
Therefore in any case,
\beq \prob{(\hat{\Ybf}_{B_2},\xbf'_{A_1},\xbf'_{A_2})\in
T_{\delta}} \leq 2^{-TI(\hat{Y}_{B_2};X_{A_1}|X_{A_2})}, \eeq
Similarly we can show that,
\beq \prob{(\hat{\Ybf}_{D},\xbf'_{B_1},\xbf'_{B_2})\in
T_{\delta}} \leq 2^{-TI(\hat{Y}_{D};X_{B_1}|X_{B_2})}, \eeq
Therefore for any typical sequence $\xbf_{\mathcal{V}}'$, the probability that  (\ref{eq:ExmCond1})-(\ref{eq:ExmCond3}) are satisfied is upper bounded by
\begin{align} \nonumber & 2^{-TI(X_S;Y_{A_2})} \times 2^{-TI(\hat{Y}_{B_2};X_{A_1}|X_{A_2})}\times 2^{-TI(\hat{Y}_{D};X_{B_1}|X_{B_2})} \\
& = 2^{-TI(X_{\Omega};\hat{Y}_{\Omega^c}|X_{\Omega^c})}
\end{align}
Now, by using the union bound over all possible $\xbf_{\mathcal{V}}'$'s and cuts, the probability of confusing $w$ with $w'$ can be bounded by
\beq \prob{w \rightarrow w'} \leq |\mathcal{X}_{\mathcal{V}}(w')| \sum_{\Omega}  2^{-T I \lp  X_{\Omega}; \hat{Y}_{\Omega^c} |X_{\Omega^c} \rp } \eeq
In the next section, we make these arguments precise, and by bounding $|\mathcal{X}_{\mathcal{V}}(w')| $ we  prove our main theorem \ref{thm:Main} for networks with a layered structure.

\subsection{Proof for layered networks}
\label{subsec:ProofLay}
In this section we extend the idea from section \ref{subsec:PfIdea} and analyze a $l_D$-layer network, $\mathcal{G}$.

Based on the proof strategy illustrated in section \ref{subsec:PfIdea}, we proceed with the error probability analysis of our scheme that was described in section \ref{subsec:EncLay}. Assume message $w$ is being transmitted. To bound the probability of error, we just need to analyze the probability that $\hat{\ybf}_D$ is jointly typical with {\em both} $w,w'$, for a message $w'$ independent of $w$. We denote this event by $w\rightarrow w'$. 

%then we have

%\beq \mathcal{P}_{\text{error}} = \prob{(\hat{\ybf}_D,w) \notin T_{\delta}}+ \prob{\exists w' \neq w ~ \text{s.t.} (\hat{\ybf}_D,w') \in T_{\delta} ~\&~ (\hat{\ybf}_D,w) \in T_{\delta}} \eeq
%By law of large numbers, the probability of the first event becomes arbitrarily small as communication block length, $T$, goes to infinity. So we just need to analyze the probability that $\hat{\ybf}_D$ is jointly
%typical with {\em both} $w,w'$ for $w'$ independent of $w$. We denote this event by $w\rightarrow w'$.

If  $\hat{\ybf}_D$ is jointly typical with $w'$, then there must be a  typical transmit sequence $\xbf_{\mathcal{V}}' \in \mathcal{X}_{\mathcal{V}}(w')$ under $w'$ such that $(\hat{\Ybf}_{D},\xbf_{\gamma_{l_D-1}}') \in T_{\delta} $, where $\gamma_{l_D-1}$ is the set of nodes at layer $l_D-1$ of the network. This means that the destination thinks this is a plausible sequence. Therefore, there is a natural source-destinationcut, $\Omega$, in $\mathcal{G}$ such that the nodes on the right hand side of the cut ({\em i.e.}  in $\Omega$) can tell $\xbf_{\mathcal{V}}'$ is not a plausible sequence, and those on the left hand side of the cut ({\em i.e.}  in $\Omega^c$) can not.  Note that due to the layered structure of the network, for any such cut, $\Omega$, we can create $d=l_D$ disjoint sub-networks of nodes corresponding to each layer of the network, with $\beta_{l-1}(\Omega)$ nodes at distance $l-1$ from $S$ that are in $\Omega$, on one side and $\beta_l(\Omega^c)$ nodes at distance $l$ from $S$ that are in $\Omega^c$, on the other, for
$l=1,\ldots,l_D$. Hence, by definition we have
\beq (\hat{\Ybf}_{\beta_l(\Omega^c)},\xbf_{\beta_{l-1}(\Omega)}',\xbf_{\beta_{l-1}(\Omega^c)}') \in T_{\delta} , \quad l=1,\ldots, l_D\eeq

Therefore, similar to the pairwise error analysis done in section \ref{subsec:PfIdea}, we can show
\beq
\label{eq:errorPair} \prob{w \rightarrow w'} \leq   
  | \mathcal{X}_{\mathcal{V}}(w')| \sum_{\Omega}   2^{-T I \lp  X_{\Omega}; \hat{Y}_{\Omega^c} |X_{\Omega^c} \rp }\eeq

%
%As the last ingredient of the proof, we state the following lemma whose proof is in appendix \ref{app:conEntropy}.

As the last ingredient of the proof, we state the following lemma which is proved in the appendix.
\begin{lemma} \label{lem:condEntropySet} Consider a layered Gaussian relay network, $\mathcal{G}$, then,
\beq  | \mathcal{X}_{\mathcal{V}}(w')|  \leq 2^{T \kappa_1} \eeq where $\kappa_1=|\mathcal{V}| $ is a constant depending on the total number of nodes in $\mathcal{G}$.
\end{lemma}

%Now by (\ref{eq:errorPair}) and lemma \ref{lem:condEntropySet}, we have
%\begin{eqnarray}
%\mathbb{P}_e & \leq & 2^{RT}  \prob{w \rightarrow w'} \\
%& \leq & 2^{RT}\sum_{\Omega} 2^{-T( I \lp  X_{\Omega}; \hat{Y}_{\Omega^c} |X_{\Omega^c} \rp -\kappa_1)}
%\end{eqnarray}

\begin{figure*} [htp]
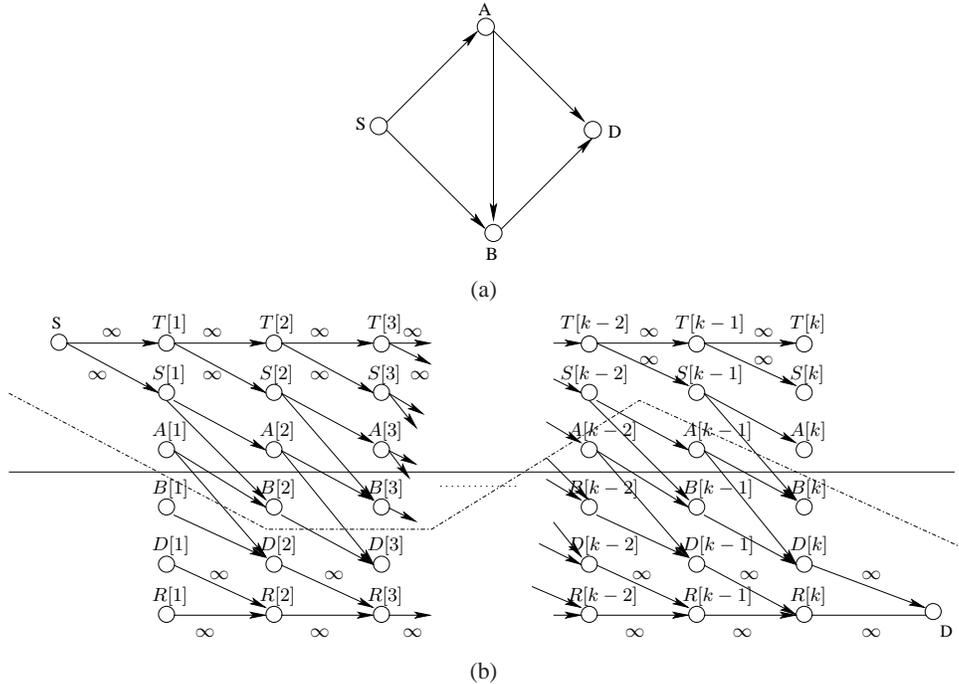

 %    \centering \subfigure[An example of general Gaussian network ]{
    \centering \subfigure[]{
       \scalebox{0.8}{ \input{unfolding_a_2.pstex_t} } 
}
     \hspace{.4in}
     \subfigure[]{
       \scalebox{0.8}{ \input{unfolding_b_2.pstex_t} } 
}
\caption{An example of a general Gaussian network with un equal
paths from S to D is shown in $(a)$.  The corresponding unfolded
network is shown in $(b)$. An example of steady cuts and wiggling cuts are respectively shown in $(b)$ by solid and dotted lines.}
\label{fig:unfolding}
\end{figure*}

Therefore, by (\ref{eq:errorPair}) and lemma \ref{lem:condEntropySet}, we have the following,
\begin{lemma} \label{lem:MainLay} Given a Gaussian relay network $\mathcal{G}$ with a layered structure, all rates $R$ satisfying the following condition are achievable,
\beq 
R <  \min_{\Omega\in\Lambda_D}
I(\hat{Y}_{\Omega^c};X_{\Omega}|X_{\Omega^c})-\kappa_1 \eeq
where $X_i$, $i \in \mathcal{V}$, are iid with complex normal (Gaussian) distribution, and $\kappa_1=|\mathcal{V}|$ is a constant depending on the total number of nodes in $\mathcal{G}$.
\end{lemma}

To prove our main theorem \ref{thm:Main} for layered networks, we state the following lemma which is proved in the appendix, 
\begin{lemma} \label{lem:BeamForming}
Given a Gaussian relay network $\mathcal{G}$, then
\beq \overline{C}-\min_{\Omega\in\Lambda_D}
I(\hat{Y}_{\Omega^c};X_{\Omega}|X_{\Omega^c})< \kappa_2 \eeq
where $X_i$, $i \in \mathcal{V}$, are iid with complex normal (Gaussian) distribution, $\overline{C}$ is the cut-set upper bound on the capacity of $\mathcal{G}$ as described in equation (\ref{eq:MinCut}), and $\kappa_2=2 |\mathcal{V}| $.
\end{lemma}

Now by lemma \ref{lem:MainLay} and lemma \ref{lem:BeamForming}, we have the following main result
\begin{theorem} \label{thm:MainLay} Given a Gaussian relay network $\mathcal{G}$ with a layered structure, all rates $R$ satisfying the following condition are achievable,
\beq 
R <  \overline{C}-\kappa_{\text{Lay}} \eeq
where $\overline{C}$ is the cut-set upper bound on the capacity of $\mathcal{G}$ as described in equation (\ref{eq:MinCut}), and $\kappa_{\text{Lay}}=\kappa_1+\kappa_2=3 |\mathcal{V}| $ is a constant depending on the total number of nodes in $\mathcal{G}$ (denoted by $|\mathcal{V}|$).
\end{theorem}

\section{Proof for general networks}
\label{sec:Gen}

Given the proof for layered networks with equal path lengths, we are
ready to tackle the proof of Theorem \ref{thm:Main}  for general Gaussian relay networks.

The ingredients are developed below. First is that any Gaussian network can be
unfolded over time to create a layered  Gaussian network (this
idea was introduced for graphs in \cite{ACLY00} to handle cycles in a
graph). The idea is to unfold the network to $K$ stages such that i-th
stage is representing what happens in the network during $(i-1)T$ to
$iT-1$ symbol times. For example in figure \ref{fig:unfolding}(a) a
network with unequal paths from $S$ to $D$ is shown. Figure
\ref{fig:unfolding}(b) shows the unfolded form of this network. As we
notice each node $V \in \mathcal{V}$ is appearing at stage $1 \leq i
\leq K$ as $V[i]$.  Now we state the following lemma which is a corollary of Theorem \ref{thm:MainLay}

%One should notice that in general there
%must be an infinite capacity link between the same node and itself
%appearing at different times however, here we are omitting these links
%which means we limit the nodes to have a finite memory $T$. Now we
%state the following lemma which is a corollary of theorem \ref{thm:MainLay}

\begin{lemma}
\label{lem:UnfGen}
Given a Gaussian relay network, $\mathcal{G}$, all rates $R$ satisfying the following condition are achievable, 
\begin{eqnarray}
\label{eq:GenNetAchiRate}
R < \frac{1}{K}  \min_{\Omega_{\text{unf}}\in\Lambda_D}
I(Y_{\Omega_{\text{unf}}^c};X_{\Omega_{\text{unf}}} |X_{\Omega_{\text{unf}}^c}) - \kappa_1
\end{eqnarray}
where $\mathcal{G}_{\text{unf}}^{(K)}$ is the time expanded graph associated with $\mathcal{G}$, random variables $\{X_i[t]\}_{1 \leq t \leq K}, i \in \mathcal{V}$ are iid with complex normal (Gaussian) distribution, and  $\kappa_1=3|\mathcal{V}| $.
\end{lemma}
\begin{proof}
By unfolding $\mathcal{G}$ we get an acyclic 
network such that all the paths from the source to the destination
have equal length. Therefore, by theorem \ref{thm:MainLay}, all rates $R_{\text{unf}}$,  satisfying the following condition are achievable in the time-expanded graph
\beq \displaystyle R_{\text{unf}} <
\min_{\Omega_{\text{unf}}\in\Lambda_D} I(Y_{\Omega_{\text{unf}}^c};X_{\Omega_{\text{unf}}} |X_{\Omega_{\text{unf}}^c}) - \kappa_{\text{unf}}
\eeq
where $\{X_i[t]\}_{1 \leq t \leq K}, i \in \mathcal{V}$ are iid with complex normal (Gaussian) distribution, and $\kappa_{\text{unf}}=K |\mathcal{V}| \log 4 \eta$. Since it takes $K$ steps to translate and achievable scheme in the time-expanded graph to an achievable scheme in the original graph, and $\kappa_1=\frac{1}{K}\kappa_{\text{unf}}=|\mathcal{V}| \log 4 \eta$, then the Lemma is proved. 
\end{proof}

Note that the general achievability scheme that we use here is similar to the one described in section \ref{subsec:EncLay} for layered networks, except now the message $W\in\{1,2,\ldots,2^{KRT}\}$ is encoded by the source $S$ into a signal over $KT$
transmission times (symbols). Still, each relay operates over blocks of time $T$ symbols.  In particular each received sequence $\ybf_i^{(k)}$ at node $i$ is quantized into $\hat{\ybf}_i^{(k)}$ which is then randomly mapped  into a Gaussian codeword $\xbf_i^{(k)}$ using a random (binning) function $f_i(\hat{ \ybf}_i^{(k)})$.  Given the knowledge of all the encoding functions at the relays and signals received over $K+|V|-2$ blocks, the decoder D, attempts to decode the message W sent by the source.

If we look at different cuts in the time-expanded graph we notice that
there are two types of cuts.  One type separates the nodes at
different stages identically. An example of such a steady cut is drawn
with solid line in figure \ref{fig:unfolding} (b). However there is another type of cut which does
not behave identically at different stages. An example of such a
wiggling cut is drawn with dotted line in figure \ref{fig:unfolding}
(b). There is no correspondence between these cuts and the cuts in the
original network. 

Now comparing Lemma \ref{lem:UnfGen} to the main Theorem
\ref{thm:Main} we want to prove, we notice that in this lemma the
achievable rate is found by taking the minimum of cut-values over all
cuts in the time-expanded graph (steady and wiggling ones as shown in figure \ref{fig:unfolding}). However in
theorem \ref{thm:Main} we want to prove that we can achieve a
rate by taking the minimum of cut-values over only the cuts in the
original graph or similarly over the steady cuts in the time-expanded
network. In the following lemma, which is proved in the appendix, we show  that asymptotically as $K \rightarrow \infty$ this difference (normalized by $1/K$) vanishes.

\begin{lemma}
\label{lem:trellis_min_cut} Consider a Gaussian relay network, $\mathcal{G}$. Then for any cut $\Omega_{\text{unf}}$ on the unfolded graph we have, \beq
(K-L+1)\min_{\Omega \in \Lambda_{D}}
I(Y_{\Omega^c};X_{\Omega}|X_{\Omega^c}) \leq
I(Y_{\Omega_{\text{unf}}^c};X_{\Omega_{\text{unf}}}|X_{\Omega_{\text{unf}}^c})  \eeq
where $L=2^{|\mathcal{V}|-2}$, $X_{i \in \mathcal{V}}$ are iid with complex normal (Gaussian) distribution, and
$\{X_i[t]\}_{1 \leq t \leq K}, i \in \mathcal{V}$ are also iid with complex normal (Gaussian) distribution.
\end{lemma}

Hence, by lemma \ref{lem:UnfGen} and lemma \ref{lem:trellis_min_cut} we have the following lemma,
\begin{lemma} \label{lem:MainArbit} Given a Gaussian relay network $\mathcal{G}$, all rates $R$ satisfying the following condition are achievable, 
\beq 
R <  \min_{\Omega\in\Lambda_D}
I(Y_{\Omega^c};X_{\Omega}|X_{\Omega^c})-\kappa_1 \eeq
where $X_i$, $i \in \mathcal{V}$, are i.i.d. with complex normal (Gaussian) distribution, and $\kappa_1=3 |\mathcal{V}| $.
\end{lemma}
Now by lemma \ref{lem:BeamForming} we know that,
\begin{eqnarray} \nonumber \overline{C}-\min_{\Omega\in\Lambda_D}
I(Y_{\Omega^c};X_{\Omega}|X_{\Omega^c}) & \leq & \overline{C}-\min_{\Omega\in\Lambda_D} 
I(\hat{Y}_{\Omega^c};X_{\Omega}|X_{\Omega^c}) \\ \label{eq:genSum} &\leq & 2 |\mathcal{V}| \end{eqnarray}
where $X_i$, $i \in \mathcal{V}$, are iid with complex normal (Gaussian) distribution.

Therefore, by lemma \ref{lem:MainArbit} and inequality (\ref{eq:genSum}) all rates up to $\overline{C}-|\mathcal{V}|(3 + 2)=\overline{C}-5 |\mathcal{V}|$ are achieved and the proof of our main theorem \ref{thm:Main} is complete.

%The ingredients are developed below. First is that any network can be
%unfolded over time to create a layered deterministic network (this
%idea was introduced for graphs in \cite{ACLY00} to handle cycles in a
%graph). The idea is to unfold the network to $K$ stages such that i-th
%stage is representing what happens in the network during $(i-1)T$ to
%$iT-1$ symbol times. For example in figure \ref{fig:unfolding}(a) a
%network with unequal paths from $S$ to $D$ is shown. Figure
%\ref{fig:unfolding}(b) shows the unfolded form of this network. As we
%notice each node $v \in \mathcal{V}$ is appearing at stage $1 \leq i
%\leq K$ as $v[i]$. There are additional nodes: $T[i]$'s and
%$T[i]$'s. These nodes are just virtual transmitters and receivers that
%are put to buffer and synchronize the network. Since all communication
%links connected to these nodes ($T[i]$'s and $R[i]$'s) are modelled as
%wireline links without any capacity limit they would not impose any
%constraint on the network. One should notice that in general there
%must be an infinite capacity link between the same node and itself
%appearing at different times however, here we are omitting these links
%which means we limit the nodes to have a finite memory $T$.

%%\label{sec:Disc}

\vspace{0.1in} \noindent {\bf Acknowledgements}: The
research of D.  Tse and A. Avestimehr are supported by the National
Science Foundation through grant CCR-01-18784 and the ITR grant:"The
3R's of Spectrum Management:Reuse, Reduce and Recycle.". The research
of S. Diggavi is supported in part by the Swiss National Science
Foundation NCCR-MICS center.

%\newpage

\appendices
\section{proof of beam-forming lemma}
We know that the capacity of a $r \times t$ MIMO  channel $H$, with water filling is
\beq
C_{wf}=\sum_{i=1}^n \log (1+\tilde{Q}_{ii} \lambda_i)
\eeq
where $n=min(r,t)$, and $\lambda_i$'s are the singular values of $H$ and $\tilde{Q}_{ii}$ is given by water filling solution satisfying 
\beq \sum_{i=1}^n \tilde{Q}_{ii}=nP \eeq

With equal power allocation
\beq C_{ep}=\sum_{i=1}^n \log (1+P \lambda_i) \eeq
Now note that
\begin{eqnarray}
C_{wf}-C_{ep}&=& \log \lp \frac{\prod_{i=1}^n  (1+\tilde{Q}_{ii} \lambda_i)}{\prod_{i=1}^n  (1+P \lambda_i)}\rp  \\
& \leq & \log \lp \frac{\prod_{i=1}^n  (1+\tilde{Q}_{ii} \lambda_i)}{\prod_{i=1}^n  \max (1,P \lambda_i)}\rp \\
& = & \log \lp \prod_{i=1}^n \frac{1+\tilde{Q}_{ii} \lambda_i}{ \max (1,P \lambda_i)}\rp \\
& = & \log \lp \prod_{i=1}^n \lp \frac{1}{ \max (1,P \lambda_i)}+ \frac{\tilde{Q}_{ii} \lambda_i}{ \max (1,P \lambda_i)} \rp \rp \\
& \leq & \log \lp \prod_{i=1}^n \lp 1+ \frac{\tilde{Q}_{ii} \lambda_i}{P \lambda_i} \rp \rp \\
& = & \log \lp \prod_{i=1}^n \lp 1+ \frac{\tilde{Q}_{ii} }{P} \rp \rp 
\end{eqnarray}
Now note that 
\beq \sum_{i=1}^n (1+ \frac{\tilde{Q}_{ii} }{P})=2n \eeq and therefore by arithmetic mean-geometric mean inequality we have
\beq \prod_{i=1}^n \lp 1+ \frac{\tilde{Q}_{ii} }{P} \rp \leq \lp \frac{ \sum_{i=1}^n (1+ \frac{\tilde{Q}_{ii} }{P})}{n} \rp ^n=2^n \eeq
and hence 
\beq  C_{ep}-C_{wf} \leq n \eeq
Hence,
\beq \label{eq:beamFormingGap} \overline{C} \leq \min_{\Omega\in\Lambda_D}
I(Y_{\Omega^c};X_{\Omega}|X_{\Omega^c}) + |\mathcal{V}|  \eeq
where $X_i$, $i \in \mathcal{V}$, are restricted to be iid with complex normal (Gaussian) distribution.

Next note that $\hat{Y}$ is obtained by quantizing $Y$ at the noise level. The effect of quantization noise can be compensated by adding a factor of two more power at each transmitter. Therefore, for each cut $\Omega$ we have
\beq \label{eq:quantizationGap} I(Y_{\Omega^c};X_{\Omega}|X_{\Omega^c})  \leq I(\hat{Y}_{\Omega^c};X_{\Omega}|X_{\Omega^c}) + |\mathcal{V}|\log 2  \eeq where $X_i$, $i \in \mathcal{V}$, are restricted to be iid with complex normal (Gaussian) distribution. Now by (\ref{eq:beamFormingGap}) and (\ref{eq:quantizationGap}), the lemma is proved.

%\appendices 
\section{proof of lemma \ref{lem:condEntropySet}}  \label{app:conEntropy} 

Assume message $w'$ is transmitted. Consider a relay, $R$, at the first layer. Then, the total number of quantized outputs at $R$ would be
\beq 2^{H(\hat{\Ybf}_R|\Xbf_S)}=2^{TI(Y_R;\hat{Y}_R|X_S)}  \eeq
Since we are using an optimal Gaussian vector quantizer at the noise level (i.e. with distortion 1), we can write
\beq \hat{Y}_R=\alpha Y_R+ N, \eeq
where $N \sim \mathcal{CN}(0,\sigma^2_N)$ is a complex Gaussian noise independent of $Y_R$ and 
\beq
\alpha = \frac{\sigma^2_Y-1}{\sigma^2_Y} ,\quad \sigma^2_N=(1-\alpha^2)\sigma^2_Y-1
\eeq
Hence
\begin{eqnarray} I(Y_R;\hat{Y}_R|X_S) &=& \log \lp 1+\frac{\alpha^2}{\sigma^2_N} \rp \\
&=& \log (1+\alpha)  \leq  1
\end{eqnarray}
Hence the list size of $R$ would be smaller than $2^T$. Now the list of typical transmit sequences can be viewed as a tree such that at each node, due to the noise, each path will be branched to at most $2^T$ other typical possibilities. Therefore, the total number of typical transmit sequences would be smaller than the product of the expansion coefficient ({\em i.e.} $2^T$) over all nodes in the graph. Or, more precisely
\begin{align}
& \log \lp |\mathcal{X}_{\mathcal{V}}(w')| \rp =\log \lp |\mathcal{Y}_{\mathcal{V}}(w')| \rp\\
& \quad = H(\hat{\Ybf}_{\mathcal{V}}|w')  = \sum_{l=1}^{l_D} H(\hat{\Ybf}_{\gamma_{l}}|\hat{\Ybf}_{\gamma_{l-1}}) \\
& \quad = \sum_{l=1}^{l_D} H(\hat{\Ybf}_{\gamma_{l}}|\Xbf_{\gamma_{l-1}}) \\
& \quad \leq \sum_{l=1}^{l_D} T |\gamma_{l}|= T |\mathcal{V}|
\end{align}
Where $\gamma_l$ is the set of nodes at the $l$-th layer of the network.
Hence,
\beq |\mathcal{X}_{\mathcal{V}}(w')| \leq 2^{T|\mathcal{V}|} \eeq and the proof is complete.

\section{proof of lemma \ref{lem:trellis_min_cut}}  \label{app:submodularity} 
First, we prove a lemma which is a slight generalization of lemma 6.4 in \cite{ADTallerton2_07},

\begin{lemma} \label{lem:loop}
Let $\mathcal{V}_1,\ldots,\mathcal{V}_l$ be $l$ non identical subsets of
$\mathcal{V}-\{S\}$ such that $D \in \mathcal{V}_i$ for all $1 \leq i \leq
l$. Also assume a product distribution on continuous random variables $X_i,~i \in \mathcal{V}$.
Then
{\footnotesize
\beq h(Y_{\mathcal{V}_{2}}|X_{\mathcal{V}_1})+\cdots
+h(Y_{\mathcal{V}_{l}}|X_{\mathcal{V}_{l-1}})+h(Y_{\mathcal{V}_{1}}|X_{\mathcal{V}_l}) \geq
\sum_{i=1}^l H(Y_{\tilde{\mathcal{V}}_i}|X_{\tilde{\mathcal{V}}_i}) \eeq } where for $k=1,\ldots,l$,
\begin{eqnarray}
\tilde{\mathcal{V}}_{k}&=&\bigcup_{\{i_1,\ldots,i_{k} \} \subseteq
\{1,\ldots,l \}} (\mathcal{V}_{i_1}\cap \cdots \cap \mathcal{V}_{i_{k}} )
\end{eqnarray}
\iffalse
\begin{eqnarray}
\tilde{\mathcal{V}}_1&=& \bigcup_{i_1=1}^l \mathcal{V}_{i_1}\\ \tilde{\mathcal{V}}_2&=&
\bigcup_{\{i_1,i_2 \} \subseteq \{1,\ldots,l \}} ( \mathcal{V}_{i_1}\cap \mathcal{V}_{i_2}
) \\ \nonumber \vdots && \\
\tilde{\mathcal{V}}_{k-1}&=&\bigcup_{\{i_1,\ldots,i_{l-1} \} \subseteq
\{1,\ldots,l \}} (\mathcal{V}_{i_1}\cap \cdots \cap \mathcal{V}_{i_{l-1}} ) \\ \tilde{\mathcal{V}}_l
&=& \mathcal{V}_{1}\cap \cdots \cap \mathcal{V}_{l}
\end{eqnarray}
\fi
or in another words each $\tilde{\mathcal{V}}_j$ is the union of $l \choose
j$ sets such that each set is intersect of $j$ of $\mathcal{V}_i$'s.
\end{lemma}

\begin{proof}
First note that
\begin{scriptsize}
\begin{eqnarray}
\nonumber &&
h(Y_{\mathcal{V}_2}|X_{\mathcal{V}_1})+\cdots+h(Y_{\mathcal{V}_l}|X_{\mathcal{V}_{l-1}})+h(Y_{\mathcal{V}_1}|X_{\mathcal{V}_l})=
\\ && \nonumber
h(Y_{\mathcal{V}_2},X_{\mathcal{V}_1})+\cdots+h(Y_{\mathcal{V}_l},X_{\mathcal{V}_{l-1}})+h(Y_{\mathcal{V}_1},X_{\mathcal{V}_l})
- \sum_{i=1}^l h(X_{\mathcal{V}_i}) \end{eqnarray}
\end{scriptsize}
and

\begin{eqnarray} \sum_{i=1}^l h(Y_{\tilde{\mathcal{V}}_i}|X_{\tilde{\mathcal{V}}_i}) &=& \sum_{i=1}^l h(Y_{\tilde{\mathcal{V}}_i},X_{\tilde{\mathcal{V}}_i})- \sum_{i=1}^l
 h(X_{\tilde{\mathcal{V}}_i}) \end{eqnarray}

Now define the set \beq \mathcal{W}_i=\{ Y_{\mathcal{V}_i},X_{\mathcal{V}_{i-1}}\}, \quad i=
1,\ldots,l \eeq where $\mathcal{V}_0=\mathcal{V}_l$.  

It  is easy to show that,
\beq \sum_{i=1}^l h(X_{\mathcal{V}_i}) =
\sum_{i=1}^l h(X_{\tilde{\mathcal{V}}_i}) \eeq Therefore, we just need to prove that \beq
\sum_{i=1}^l h(\mathcal{W}_i) \geq \sum_{i=1}^l
h(Y_{\tilde{\mathcal{V}}_i},X_{\tilde{\mathcal{V}}_i}) \eeq 

Now, since the differential entropy function is a
submodular function we have, \beq \label{eqn:use_k_way} \sum_{i=1}^l h(\mathcal{W}_i)
\geq \sum_{i=1}^l h(\tilde{\mathcal{W}}_i)\eeq where \beq
\tilde{\mathcal{W}}_r=\bigcup_{\{i_1,\ldots,i_r\} \subseteq \{1,\ldots,l \}}
(\mathcal{W}_{i_1} \cap \cdots \cap \mathcal{W}_{i_r} ), \quad r=1,\ldots,l \eeq Now for
any $r$ ($1 \leq r \leq l$) we have

\begin{align*}
\displaystyle
&\tilde{\mathcal{W}}_r=\bigcup_{\{i_1,\ldots,i_r\} \subseteq \{1,\ldots,l \}}
(\mathcal{W}_{i_1} \cap \cdots \cap \mathcal{W}_{i_r}) \\ &= \bigcup_{\{i_1,\ldots,i_r\}
\subseteq \{1,\ldots,l \}} (\{ Y_{\mathcal{V}_{i_1}},X_{\mathcal{V}_{i_1-1}} \} \cap
\cdots \cap \{ Y_{\mathcal{V}_{i_r}}X_{\mathcal{V}_{i_r-1}} \}) \\ &=
\bigcup_{\{i_1,\ldots,i_r\} \subseteq \{1,\ldots,l \}} (\{ Y_{\mathcal{V}_{i_1}
\cap \cdots \cap \mathcal{V}_{i_r} },X_{\mathcal{V}_{(i_1-1)} \cap \cdots \cap
X_{\mathcal{V}_{(i_r-1)}}} \}) \\ \nonumber &= \left \{
Y_{\bigcup_{\{i_1,\ldots,i_r\} } (\mathcal{V}_{i_1} \cap \cdots \cap \mathcal{V}_{i_r})
},X_{\bigcup_{\{i_1,\ldots,i_r\} }(\mathcal{V}_{(i_1-1)} \cap \cdots \cap
\mathcal{V}_{(i_r-1)})} \right \} \\ &= \{Y_{\tilde{\mathcal{V}}_r},X_{\tilde{\mathcal{V}}_r} \}
\end{align*} 
Therefore by equation (\ref{eqn:use_k_way}) we have,
\begin{eqnarray} \sum_{i=1}^l h(\mathcal{W}_i)
& \geq & \sum_{i=1}^l h(\tilde{\mathcal{W}}_i) \\
&=&\sum_{i=1}^l h(Y_{\tilde{\mathcal{V}}_i},X_{\tilde{\mathcal{V}}_i})
\end{eqnarray}

Hence the Lemma is proved.

\end{proof}

Now we are ready to prove lemma \ref{lem:trellis_min_cut}.
First note that any cut in the unfolded graph, $\Omega_{\text{unf}}$, partitions
the nodes at each stage $1\leq i \leq K$ to $\mathcal{U}_i$ (on the
left of the cut) and $\mathcal{V}_i$ (on the right of the cut). If at
one stage $S[i] \in \mathcal{V}_i$ or $D[i] \in \mathcal{U}_i$ then
the cut passes through one of the infinite capacity edges (capacity
$Kq$) and hence the lemma is obviously
proved. Therefore without loss of generality assume that $S[i] \in
\mathcal{U}_i$ and $D[i] \in \mathcal{V}_i$ for all $1 \leq i \leq
K$. Now since for each $i \in \mathcal{V}$, $\{x_i[t]\}_{1 \leq t \leq
K}$ are i.i.d distributed we can write

\begin{eqnarray}
I(Y_{\Omega_{\text{unf}}^c};X_{\Omega_{\text{unf}}}|X_{\Omega_{\text{unf}}^c})&=&
\sum_{i=1}^{K-1} I(Y_{\mathcal{V}_{i+1}};X_{\mathcal{U}_i}|X_{\mathcal{V}_i}) 
\end{eqnarray}

Consider the sequence of $\mathcal{V}_i$'s. Note that there are total of
$L=2^{|\mathcal{V}|-2}$ possible subsets of $\mathcal{V}$ that contain $D$ but not
$S$. Assume that $\mathcal{V}_s$ is the first set that is revisited. Assume that
it is revisited at step $\mathcal{V}_{s+l}$. We have,
{\footnotesize
\beq  \sum_{i=s}^{s+l-1} I(Y_{\mathcal{V}_{i+1}};X_{\mathcal{U}_i}|X_{\mathcal{V}_i})=
 \sum_{i=s}^{s+l-1} h(Y_{\mathcal{V}_{i+1}}|X_{\mathcal{V}_i}) -  h(Y_{\mathcal{V}_{i+1}}|X_{\mathcal{V}_i},X_{\mathcal{U}_i}) \eeq
}

Now by Lemma \ref{lem:loop} 
we have \beq \label{eq:submodularity} \sum_{i=s}^{s+l-1} h(Y_{\mathcal{V}_{i+1}}|X_{\mathcal{V}_i}) \geq \sum_{i=1}^l
h(Y_{\tilde{\mathcal{V}}_i}|X_{\tilde{\mathcal{V}}_i})  \eeq where $\tilde{\mathcal{V}}_i$'s are as described
in lemma \ref{lem:loop}. Next, note that  $h(Y_{\mathcal{V}_{i+1}}|X_{\mathcal{V}_i},X_{\mathcal{U}_i})$ is just the entropy of channel noises, and since for any $v \in \mathcal{V}$ we have \beq |\{i|v \in
\mathcal{V}_i\}|=|\{j|v \in \tilde{\mathcal{V}}_j\}|\eeq
, we get
\beq \label{eq:noise_entropy} \sum_{i=s}^{s+l-1} h(Y_{\mathcal{V}_{i+1}}|X_{\mathcal{V}_i},X_{\mathcal{U}_i})= \sum_{i=1}^l
h(Y_{\tilde{\mathcal{V}}_i}|X_{\tilde{\mathcal{V}}_i},X_{\tilde{\mathcal{V}}_i}^c) \eeq
Now by putting (\ref{eq:submodularity}) and (\ref{eq:noise_entropy}) together, we get
\begin{eqnarray} \sum_{i=s}^{s+l-1} I(Y_{\mathcal{V}_{i+1}};X_{\mathcal{U}_i}|X_{\mathcal{V}_i}) & \geq & \sum_{i=1}^l
I(Y_{\tilde{\mathcal{V}}_i};X_{\tilde{\mathcal{V}}_i^c}|X_{\tilde{\mathcal{V}}_i})\\
& \geq & l \min_{\Omega \in \Lambda_{D}}
I(Y_{\Omega^c};X_{\Omega}|X_{\Omega^c}) \quad ~~~  \end{eqnarray}
 Now since in any $L-1$ time frame there is at least one loop, therefore except at most a path of length $L-1$
everything in $\sum_{i=1}^{K-1} I(Y_{\mathcal{V}_{i+1}};X_{\mathcal{U}_i}|X_{\mathcal{V}_i})  $. can be replaced with the value of the min-cut. Therefore, \beq \sum_{i=1}^{K-1}
I(Y_{\mathcal{V}_{i+1}};X_{\mathcal{U}_i}|X_{\mathcal{V}_i}) \geq (K-L+1)\min_{\Omega \in \Lambda_{D}}
I(Y_{\Omega^c};X_{\Omega}|X_{\Omega^c}) \eeq
and hence the proof is complete.

\end{document}